\documentclass[12pt]{article}

\makeatletter

\def\paragraph{\@startsection{paragraph}{4}{\z@}{+2.00ex plus
 +1ex minus +.2ex}{0.5ex plus .2ex}{\it\normalsize}}

\def\section{\@startsection {section}{1}{\z@}{+3.0ex plus +1ex minus
  +.2ex}{2.3ex plus .2ex}{\normalsize\bf\boldmath}}
\def\subsection{\@startsection{subsection}{2}{\z@}{+2.5ex plus +1ex
minus +.2ex}{1.5ex plus .2ex}{\normalsize\bf\boldmath}}
\def\subsubsection{\@startsection{subsubsection}{3}{\z@}{+3.25ex plus
 +1ex minus +.2ex}{1.5ex plus .2ex}{\normalsize\it}}

\expandafter\ifx\csname mathrm\endcsname\relax\def\mathrm#1{{\rm #1}}\fi

\newcounter{saveeqn}

\@addtoreset{equation}{section}

\newcount\@tempcntc
\def\@citex[#1]#2{\if@filesw\immediate\write\@auxout{\string\citation{#2}}\fi
  \@tempcnta\z@\@tempcntb\m@ne\def\@citea{}\@cite{\@for\@citeb:=#2\do
    {\@ifundefined
       {b@\@citeb}{\@citeo\@tempcntb\m@ne\@citea
        \def\@citea{,\penalty\@m\ }{\bf ?}\@warning
       {Citation `\@citeb' on page \thepage \space undefined}}%
    {\setbox\z@\hbox{\global\@tempcntc0\csname
b@\@citeb\endcsname\relax}%
     \ifnum\@tempcntc=\z@ \@citeo\@tempcntb\m@ne
       \@citea\def\@citea{,\penalty\@m}
       \hbox{\csname b@\@citeb\endcsname}%
     \else
      \advance\@tempcntb\@ne
      \ifnum\@tempcntb=\@tempcntc
      \else\advance\@tempcntb\m@ne\@citeo
      \@tempcnta\@tempcntc\@tempcntb\@tempcntc\fi\fi}}\@citeo}{#1}}

\def\@citeo{\ifnum\@tempcnta>\@tempcntb\else\@citea
  \def\@citea{,\penalty\@m}%
  \ifnum\@tempcnta=\@tempcntb\the\@tempcnta\else
   {\advance\@tempcnta\@ne\ifnum\@tempcnta=\@tempcntb \else
\def\@citea{--}\fi
    \advance\@tempcnta\m@ne\the\@tempcnta\@citea\the\@tempcntb}\fi\fi}

\def\asymp#1%
{\mathrel{\raisebox{-.4em}{$\widetilde{\scriptstyle #1}$}}}

\def\Nequal#1%
{\mathrel{\raisebox{-.5em}{$\stackrel{=}{\scriptstyle\rm#1}$}}}
\newcommand{\dsl}[1]{\not \hspace{-0.7mm}#1}
\def\dsl{\mathpalette\make@slash}
\def\make@slash#1#2{\setbox\z@\hbox{$#1#2$}%
  \hbox to 0pt{\hss$#1/$\hss\kern-\wd0}\box0}

\def\refeq#1{\mbox{(\ref{#1})}}

\def\citere#1{\mbox{Ref.~\cite{#1}}}

\newcommand{\ri}{{\mathrm{i}}}
\newcommand{\rd}{{\mathrm{d}}}

\def\eg{e.g.\ }

\def\draftdate{\relax}
\def\mda{\relax}
\def\mua{\relax}
\def\mla{\relax}
\def\draft{
\def\thtystars{******************************}
\def\sixtystars{\thtystars\thtystars}
\typeout{}
\typeout{\sixtystars**}
\typeout{* Draft mode!
         For final version remove \protect\draft\space in source file *}
\typeout{\sixtystars**}
\typeout{}
\def\draftdate{\today}
\def\mua{\marginpar[\boldmath\hfil$\uparrow$]%
                   {\boldmath$\uparrow$\hfil}%
                    \typeout{marginpar: $\uparrow$}\ignorespaces}
\def\mda{\marginpar[\boldmath\hfil$\downarrow$]%
                   {\boldmath$\downarrow$\hfil}%
                    \typeout{marginpar: $\downarrow$}\ignorespaces}
\def\mla{\marginpar[\boldmath\hfil$\rightarrow$]%
                   {\boldmath$\leftarrow $\hfil}%
                    \typeout{marginpar: $\leftrightarrow$}\ignorespaces}
\def\Mua{\marginpar[\boldmath\hfil$\Uparrow$]%
                   {\boldmath$\Uparrow$\hfil}%
                    \typeout{marginpar: $\uparrow$}\ignorespaces}
\def\Mda{\marginpar[\boldmath\hfil$\Downarrow$]%
                   {\boldmath$\Downarrow$\hfil}%
                    \typeout{marginpar: $\downarrow$}\ignorespaces}
\def\Mla{\marginpar[\boldmath\hfil$\Rightarrow$]%
                   {\boldmath$\Leftarrow $\hfil}%
                    \typeout{marginpar: $\leftrightarrow$}\ignorespaces}
\overfullrule 5pt
\oddsidemargin -15mm
\marginparwidth 29mm
}

\def\stars{\strut\leaders\hbox{*}\hfill\strut}
\def\starline{\hfil\strut\hfil\hbox to \textwidth {\stars}\hfil}

\newcommand{\W}{{\cal W}}

\renewcommand{\S}{{\cal S}}
\newcommand{\s}{{\mathbf s}}

\begin{document}
\thispagestyle{empty}
\def\thefootnote{\fnsymbol{footnote}}
\setcounter{footnote}{1}
\null
\draftdate\hfill KA-TP-19-2002 \\
\strut\hfill hep-ph/0211150 
\vfill
\begin{center}
{\Large \bf\boldmath
On the Renormalization of the \\
Minimal Supersymmetric Standard Model \par} \vskip 2em
\vspace{1cm}
{\large
{\sc Elisabeth Kraus$^1$, Markus Roth$^{2}$%
\footnote{Talk given at the 10th International Conference on
Supersymmetry and Unification of Fundamental Interactions, 
SUSY 2002 (June 17-23, 2002, DESY, Hamburg)}
 and Dominik St\"ockinger$^{3}$} } 
\\[.5cm]
$^1$ {\it Institut f\"ur Theoretische Physik, Universit\"at Bonn\\
D-53115 Bonn, Germany}
\\[0.3cm]
$^2$ {\it Institut f\"ur Theoretische Physik, Universit\"at Karlsruhe\\
D-76131 Karlsruhe, Germany}
\\[0.3cm]
$^3$ {\it Deutsches Elektron-Synchrotron DESY \\
D-22603 Hamburg, Germany}
\end{center}
{\bf Abstract:} \par 
The renormalization of the Minimal Supersymmetric Standard
Model is discussed. In particular we focus on the
soft-supersymmetry breaking sector of the MSSM and comment on 
non-renormalization theorems.
\par
\vskip 1cm
\vfill
\noindent
November 2002   
\null
\setcounter{page}{0}
\clearpage
\def\thefootnote{\arabic{footnote}}
\setcounter{footnote}{0}

\begin{center}
{\Large \bf\boldmath
On the Renormalization of the \\
Minimal Supersymmetric Standard Model \par} \vskip 2em
\vspace{0.3cm}
{\large 
{\sc Elisabeth Kraus$^1$, \underline{Markus Roth}$^{2}$ and Dominik St\"ockinger$^{3}$} } 
\\[0.5cm]
$^1$ {\it Institut f\"ur Theoretische Physik, Universit\"at Bonn,
D-53115 Bonn, Germany}
\\[0.1cm]
$^2$ {\it Institut f\"ur Theoretische Physik, Universit\"at Karlsruhe,
D-76131 Karlsruhe, Germany}
\\[0.1cm]
$^3$ {\it Deutsches Elektron-Synchrotron DESY, 
D-22603 Hamburg, Germany}
\end{center}\par
\vskip 1.0cm {\bf Abstract:} \par 
The renormalization of the Minimal Supersymmetric Standard
Model is discussed. In particular we focus on the
soft-supersymmetry breaking sector of the MSSM and comment on 
non-renormalization theorems.
\par
\vskip 1cm

\section*{Introduction}

The Minimal Supersymmetric Standard Model (MSSM) \cite{MSSM} is certainly the 
best motivated and conceptually most elaborated extension of the 
Standard Model (SM). Both models can be studied with very high precision 
at future $\mathrm e^+ \mathrm e^-$ colliders \cite{LinearCollider}. 
To account for the high experimental accuracy the full ${\cal O}(\alpha)$
corrections for most particle reactions and even the 
full ${\cal O}(\alpha^2)$ in specific cases have to be included into 
the theoretical predictions. 

However in supersymmetric theories like the MSSM, no regularization 
method is known which is mathematically consistent
 (that obeys the quantum action
principle) and preserves all symmetries.
Particularly dimensional regularization, which is mathematically consistent
in the form of \citere{DimReg}, violates supersymmetry
and chiral symmetry. For consistency such symmetry breakings have to
 be restored by adding
suitable (non-invariant) counterterms whose values are determined by the algebraic method. On the other 
hand dimensional reduction \cite{Siegel:1979wq} which is often used in 
phenomenological applications suffers from the fact that it is
mathematically not well defined \cite{Siegel:1980qs}. Explicit examples of
inconsistencies are known at the three-loop level \cite{Avdeev:1981vf} and,
best to our knowledge, even at one loop it is not yet completely proven 
that dimensional reduction conserves all symmetries. 

The restoration of symmetries is in practise an extremely complicated
and time consuming work \cite{Restoration}.
From the abstract point of view, 
 the Slavnov-Taylor (ST)
identity can be broken in the procedure of renormalization 
\begin{equation}
\S(\Gamma)=\Delta.
\end{equation}
 Owing to the quantum action 
principle, $\Delta$ 
consists of  local field polynomials with dimension $D\le 4$ and 
Faddeev-Popov charge $Q_{\Phi\Pi}=1$  
(see e.g.\ \citere{Hollik:2002mv} for details) satisfying the equation
\begin{equation}
\label{eq:consistency}
\S_\Gamma \Delta=0.
\end{equation}
If the cohomology problem 
\begin{equation}
\label{cohomology}
\S_\Gamma \hat \Delta = \Delta
\end{equation}
 can be solved, the theory 
is free of anomalies and the ST identity can be restored  by adding the 
symmetry-restoring counterterms $-\hat\Delta$ to the vertex function
$\Gamma$:
\begin{equation}
\S(\Gamma-\hat\Delta)=\S(\Gamma)-\S_\Gamma \hat\Delta=
\Delta-\S_\Gamma \hat \Delta=0.
\end{equation}

In case of the MSSM the algebraic analysis, which is the necessary
prerequisite for the  application in explicit calculations,
has been worked out in
\citere{Hollik:2002mv}. All symmetry requirements that have to be
respected
to all orders in perturbation theory have been explicitly given.
Furthermore, the symmetric 
counterterms of the MSSM have been systematically constructed, 
a complete set renormalization conditions has been defined, and
infrared finiteness has been  proven by 
power-counting arguments. 

In the following we want to concentrate on the soft supersymmetry
breaking part of the MSSM and briefly discuss non-renormalization
theorems at the end. 

\section*{Soft supersymmetry breaking}

In realistic models like the MSSM supersymmetry has to be broken
owing to the large mass difference between SM particles and their 
superpartners and to trigger the spontaneous breakdown of the 
$\mathrm{SU}(2)\times\mathrm{U}(1)$ gauge group. 
However in a quantum theory even a (softly) broken supersymmetry has 
to be maintained in a mathematically consistent way in higher orders. 
In order to fomulate symmetry identities that take into account soft
supersymmetry breaking
 we couple the soft symmetry breaking part of the MSSM 
to external fields that possess finite vacuum expectation values. 
Two kinds of external fields 
are used in the literature.

In \citere{Maggiore:1996gg} a BRS doublet $u,v$ has been used, where
the field $v$  has  a finite vacuum expecation value $v_A$,
and  the field $u$ is a Faddeev-Popov ghost. 
The special form of the BRS transformation law of BRS doublets
\begin{equation}
\label{eq:BRSdoublet}
\s u = v, \qquad \s v=0
\end{equation}
accounts for some extraordinary properties 
(see \eg \citere{PiSo}): 
\begin{itemize}
\item 
BRS doublets do not contribute to anomalies of the ST identity.
\item
Their contribution to the action can be written as 
BRS variations that fulfil the ST identity due to the nilpotency 
of the ST operator. The action and ST identity can be 
decomposed as follows (see \eg \citere{PiSo}):
\begin{eqnarray}
\Gamma&=&\Gamma|_{u,v=0}
+\int \rd^4 x \, v\frac{\delta }{\delta u} \tilde\Delta(u,v), \\
\S(\Gamma)&=&\S^0(\Gamma)
+\int \rd^4 x \, v \frac{\delta}{\delta u} \Gamma.
\end{eqnarray}
If we assume that the ST identity for $u,v=0$ is already solved
\begin{equation}
\S(\Gamma|_{u,v=0})=0,
\end{equation}
and if we add the symmetry-restoring counterterm 
$\S^0_\Gamma \tilde\Delta(u,v)$ to the action:
\begin{equation}
\label{eq:action}
\Gamma\to \Gamma=\Gamma|_{u,v=0}
+\S^0_\Gamma \tilde\Delta(u,v)
+\int \rd^4 x \, v\frac{\delta }{\delta u} \tilde\Delta
=\Gamma|_{u,v=0}+\S_\Gamma \tilde\Delta(u,v),
\end{equation}
the full ST identity is fulfilled:
\begin{equation}
\S(\Gamma)=0.
\end{equation}
\end{itemize}

Owing to the nilpotency of the ST operator, $\tilde\Delta(u,v)$ 
does not contribute to the ST identity at all. Thus, the  cohomology problem
 of the soft supersymmetry breaking part of the 
MSSM is trivial.  In the doublet approach, however, it is not possible
to distinguish physical soft breakings from unphysical
breakings. Indeed, in \citere{PiSo} an additional $R$ symmetry has to be
used to exclude the soft breaking of gauge symmetry as for example a
gauge boson mass. Then one remains with a restricted class of soft
breaking terms, which again include unphysical as well as physical
parameters. 
 Those couplings that contribute to soft supersymmetry 
breaking terms are clearly important for the calculation of observables 
since they are (at least partly) determined by the renormalization 
condition of physical mass parameters. All other 
couplings are unphysical in the sense that their renormalization
conditions do not effect physical observables. These unphysical
couplings can be savely neglected.%
\footnote{It should be noted that both the symmetry breaking $\Delta$ 
and the symmetry-restoring counterterm $\hat{\Delta}$ involve in 
general all kinds of couplings related to $u,v$, 
but in a regularization-scheme dependent way. Therefore, this dependence 
is purely artificial and does not effect physical observables.} 

In a second approach  
\cite{Hollik:2002mv,Girardello:1981wz,Hollik:2000pa} 
the soft supersymmetry breaking terms are generated from a massless 
external chiral superfield called spurion
\begin{eqnarray}
\hat{A}&=&e^{-\ri \theta \sigma^\mu \bar\theta \partial_\mu}
\left[A+\sqrt{2}\theta^\alpha a_\alpha+\theta^2 (F_A+v_A)\right],
\end{eqnarray}
where the $\theta^2$ component has an finite real vacuum 
expectation value $v_A$. In this way all soft-supersymmetry breaking 
terms of the MSSM are generated. 
For instance, the gaugino mass originates from 
(see \citere{Hollik:2002mv} for notations)  
\begin{equation}
\label{eq:gauginomass}
-\frac{1}{128 g^{\prime 2}}\int \rd^6 z \,
\hat{A} \hat{F}^{\prime \alpha} \hat{F}^\prime_\alpha
= \frac{v_A}{2} \int \rd^4 x \, \lambda^{\prime \alpha} \lambda^\prime_\alpha 
+ \mbox{spurion terms}.
\end{equation}

Since the spurion is dimensionless it can appear in arbitrary powers $\hat A^n$
in the effective action yielding an infinite number of new coupling 
constants and counterterms and, hence, it is not clear if such a
theory is
predictive at all. 
However it has been show in \citere{Hollik:2000pa} that indeed only a small
number of the spurion couplings are physically relevant and contribute
in the limit of vanishing spurion fields. 

In the following we want to show that the renormalization 
of the spurion part of the MSSM is trivial and 
can be traced back to the case of two BRS doublets. 
The spurion superfield obeys the following BRS transformations:
\begin{eqnarray}
\s A&=&
 \sqrt{2} \epsilon^\alpha a_\alpha
-\ri \xi^\mu \partial_\mu A, \\
\s a_\alpha &=&
 \sqrt{2} \epsilon_\alpha (F_A+v_A)
+\ri \sqrt{2} \sigma^\mu_{\alpha \dot{\alpha}} 
 \bar\epsilon^{\dot{\alpha}}\partial_\mu A
-\ri \xi^\mu \partial_\mu a_\alpha,\\
\s F_A&=&
 \ri \sqrt{2} \partial_\mu a^\alpha \sigma^\mu_{\alpha \dot{\alpha}} 
 \bar\epsilon^{\dot{\alpha}} 
-\ri \xi^\mu \partial_\mu F_A. 
\end{eqnarray}
If we decompose the Weyl spinor $a_\alpha$ into two auxiliary 
fields $a_1,a_2$ according to
\begin{eqnarray}
\label{eq:decomp}
a_\alpha&=&\sqrt{2} 
\left(\ri \sigma^\mu_{\alpha \dot{\alpha}} \bar\epsilon^{\dot{\alpha}} 
\partial_\mu a_1+\epsilon_\alpha a_2\right),
\end{eqnarray}
we obtain the following BRS transformations:
\begin{equation}
\begin{array}{rclcrcl}
\s a_1&=&A-\ri \xi^\mu \partial_\mu a_1, & \quad &
\s A&=&2 \ri \epsilon^\alpha \sigma^\mu_{\alpha \dot{\alpha}} 
 \bar\epsilon^{\dot{\alpha}} \partial_\mu a_1
-\ri \xi^\mu \partial_\mu A, \\
\s a_2&=&F_A+v_A-\ri \xi^\mu \partial_\mu a_2, & \quad &
\s F_A&=&2 \ri \epsilon^\alpha \sigma^\mu_{\alpha \dot{\alpha}} 
 \bar\epsilon^{\dot{\alpha}} \partial_\mu a_2
-\ri \xi^\mu \partial_\mu F_A.
\end{array}
\label{eq:brs2}
\end{equation}
We identify the two BRS doublets $u,v$ and $U,V$ by 
\begin{eqnarray}
\begin{array}[b]{rlrl}
U&=a_1,&\qquad V&=A-\ri \xi^\mu \partial_\mu a_1, \\
u&=a_2,&\qquad v&=F_A+v_A-\ri \xi^\mu \partial_\mu a_2,
\end{array}
\end{eqnarray}
which respect the usual BRS transformation law for BRS doublets.
The fields $u,v$ correspond exactly to the BRS doublet of 
\citere{Maggiore:1996gg}.
The additional fields $U,V$ are dimensionless and can appear
infinite many times in the action.   

In order to retain the superfield character of the original spurion
field we have to 
require in addition that $u,U$ has to appear in the action 
only in the combination \refeq{eq:decomp}. This implies in particular 
that $\Gamma$ depends
only on the combination $u\epsilon_\alpha$. In the approach 
of \citere{Maggiore:1996gg} such a requirement is missing and even
 with $R$ symmetry
additional interaction terms are present which to not fit in the 
superfield formulation. 

As a result, all spurion terms are BRS variations.
The physical relevant terms can be found in Appendix E of 
\citere{Hollik:2002mv}.
They have a one-to-one correspondence to the terms in the superfield 
notation (see equation (2.18) of \citere{Hollik:2002mv}). For instance the 
gaugino mass of \refeq{eq:gauginomass} corresponds to
\begin{equation}
\S_{\Gamma} \int \rd^4 x \left[
u \lambda^{\prime \alpha} \lambda^\prime_\alpha\right].
\end{equation}
Hence the cohomology problem  of the spurion part of the MSSM is as trivial
as in the case of the BRS doublet of \citere{Maggiore:1996gg}. 
Particularly only terms up to second order in the spurion
field contribute to soft-supersymmetry breaking terms and are
important for the calculation of physical observables. All other 
spurion terms can be savely ignored.

The main difference between the formulation with BRS doublets $u,v$ and
$U,V$ and those with the spurion superfield $\hat{A}$ is the actual form 
of the ST operator. In the formulation with BRS doublets the ST operator 
reads
\begin{equation}
\S(\Gamma)=\S^0(\Gamma)+\int \rd^4 x \left(V \frac{\delta}{\delta U}
+v \frac{\delta}{\delta u}
+\mbox{c.c.}\right),
\end{equation}   
while in the superfield notation it yields
\begin{equation}
\S(\Gamma)=\S^0(\Gamma)+\int \rd^4 x \left(\s A \frac{\delta}{\delta A}
+\s a^\alpha \frac{\delta}{\delta a^\alpha}
+\s F_A \frac{\delta}{\delta F_A}
+\mbox{c.c.}\right).
\end{equation}  

For practical purposes it can be more convenient to stay within the 
superfield notation. In this case it is sufficient to solve 
the ST identity in the limit $A,a_\alpha\to 0$:
\begin{equation}
\S(\Gamma)|_{A,a_\alpha=0}=0.
\end{equation}
In addition one has then to use  the $R$ Ward identity
\begin{equation}
\left.\left(\W^R \Gamma\right)\right|_{A,a_\alpha=0}=
\left.\left(\W^R \Gamma\right)\right|_{A,a_\alpha,F_A=0}
-\int \rd^4 x \left.\left(2 \ri F_A\frac{\delta \Gamma}{\delta
F_A}+\mbox{c.c.}\right)\right|_{A,a_\alpha=0}=0,
\end{equation}
yielding a relation between vertex functions including $F_A$ fields 
and those without spurion fields.

\section*{Non-renormalization theorems}

From superspace calculations \cite{superspace}
and renormalization group arguments \cite{RGsoft}
it has been known that soft supersymmetry breakings
have special renormalization properties: Up to some $D$-term contribtutions
it is possible to express divergences of soft breaking parameters in terms of
renormalization constants of the couplings and supersymmetric masses.
These improved renormalization properties escape the doublet approach as
well as the spurion approach.

Recently the non-renormalization theorems of supersymmetric field theories 
have been proven in the Wess-Zumino
gauge by extending coupling constants to external fields and
including  softly broken axial symmetry as an additional defining
symmetry into the model \cite{NRT1}. The extended model already contains the
soft breaking parameters and yields the relations between the
renormalization constants of supersymmetric parameters and of soft
breaking parameters \cite{NRT2}. In addition, the algebraic derivation 
exhibits a deep relation between two anomalies --- the axial anomaly and a
supersymmetry anomaly --- and the explicit form of the non-renormalization
theorems.

However, in this approach the symmetries and in particular the
ST identity have a more complicated form as in the spurion and doublet
approach. Hence, one could use the simpler formulation with spurion fields
for practical calculations and take the information on the
renormalization constants obtained from the abstract approach in the
extended model as an additional input and check of explicit calculations.

\end{document}